\documentclass{ws-ijmpcs}
\newcommand{\newangle}{< \hspace{-.8ex} {\scriptscriptstyle )}}

\begin{document}

\markboth{F. Bradamante}
{$h^+$, $h^{-}$ and hadron pair transverse spin asymmetries}

%
\catchline{}{}{}{}{}
%

\title{Work on the interplay among $h^+$, $h^{-}$ and hadron pair 
transverse spin asymmetries in SIDIS}

\author{Franco Bradamante\\
on behalf of the COMPASS Collaboration}

\address{Dipartimento di Fisica, Trieste University, and INFN, Sezione di 
Trieste, via A. Valerio 2\\
I-34127 Trieste, Italy\\
Franco.Bradamante@ts.infn.it}
\maketitle


\begin{abstract}
In the fragmentation of a transversely polarized quark a left-right 
asymmetry, the Collins asymmetry, is expected for each hadron produced 
in the process $\mu N \rightarrow \mu ' h^{+} h^{-} X$. 
Similarly, an asymmetry is also expected 
for the hadron pair, the dihadron asymmetry. 
Both asymmetries have been measured to be different from zero on 
transversely polarised proton targets
and have allowed for first extractions of the transversity distributions. 
From the high statistics COMPASS data we have further investigated these
asymmetries getting strong indications that the two  mechanisms are driven by 
a common physical process. 

\end{abstract}

\ccode{PACS numbers: 13.60.-r, 13.88.+e, 14.20.-c, 14.65.-q}

\par
\vspace*{1cm}
It is well known that the description of the partonic structure of the 
nucleon at leading twist requires the knowledge of three parton distribution 
functions (PDFs), the number, helicity and transversity PDFs. 
Being chirally odd, the transversity distribution is difficult to be measured 
and is still the least known of the three, but in the past ten years both the 
HERMES and the COMPASS experiments have provided unambiguous evidence that 
transversity is  different from zero~\cite{Barone:2010zz}. 
Experimentally, the transversity distribution has been probed in
 semi-inclusive deeply inelastic scattering (SIDIS) experiments on 
transversely polarized targets in two different ways. 
In the first one a target spin dependent azimuthal asymmetry in  
single-hadron production was measured, which depends on the convolution 
of transversity and a transverse-momentum dependent chirally odd 
fragmentation function, $H_1^{\perp}$, the so-called Collins 
function~\cite{Collins:1993}. 
The second process is dihadron 
production~\cite{Collins:1994ax,Jaffe:1997hf,Bianconi:1999cd}, 
where  transversity couples to a dihadron fragmentation function
$H_1^{\newangle}$. 
In both cases essential information for the extraction of transversity was 
provided by the measurements of the azimuthal asymmetries of the hadrons 
produced in $e^+ e^-$ annihilation~\cite{Seidl:2008xc,Vossen:2011fk}. 
Recently COMPASS has provided experimental evidence of  a close 
relationship between the Collins and the dihadron asymmetries, hinting 
at a common physical origin of the two fragmentation 
functions~\cite{fb_dubna,Adolph:2014fjw},
a conclusion also supported by calculations with a specific 
Monte Carlo model~\cite{Matevosyan:2013eia}.
That work has been continued and new results are here presented for the 
first time.

COMPASS (COmmon Muon and Proton Apparatus for Structure and Spectroscopy) 
is a fixed-target experiment at the CERN SPS taking data since 2002. 
The COMPASS spectrometer~\cite{Abbon:2007pq} is by now very well known 
in the scientific community and no description will be given here.
The results presented in this paper have been
extracted from the data collected with a transversely polarized proton 
(NH$_3$) target in 2010, already used to measure the
 Collins and the dihadron asymmetries\cite{Adolph:2012sn,Adolph:2014fjw}.
In the $ x$-Bjorken region where the Collins asymmetry is
different from zero and sizable ($x> 0.032$) the positive and negative 
hadron asymmetries exhibit a mirror symmetry and the dihadron asymmetry
is very close to the Collins asymmetry for positive hadrons.
This suggests 
that the anti-correlation between the azimuthal angles
of the positive and negative hadrons, 
which is present in the 
multi-hadrons fragmentation of the struck quark due to a local transverse 
momentum conservation, is also present in the Collins 
fragmentation function that describes the spin-dependent hadronization of 
a transversely polarized quark. 

\begin{figure}[bt]
  \centering
  \includegraphics[width=0.95\textwidth]{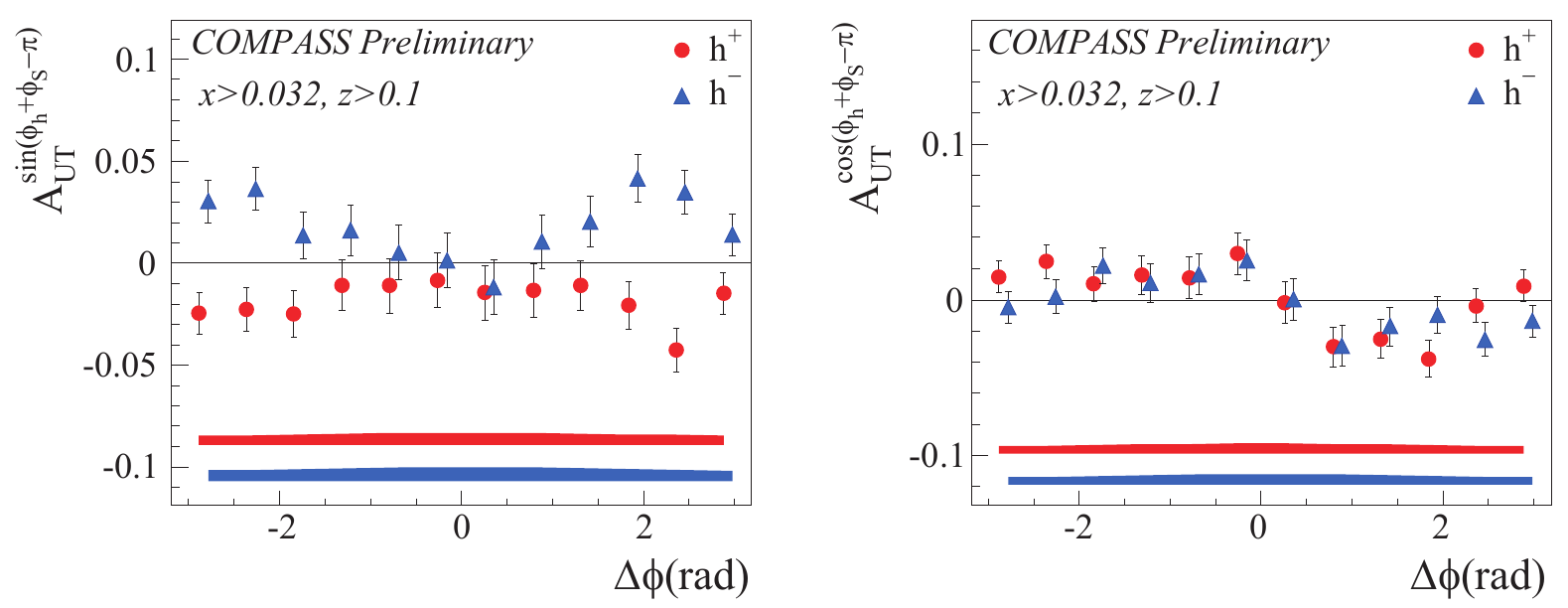}
  \caption{
Left: the $A^{\sin(\phi_1+\phi_S-\pi)}_{1CL}$ (red points) and
    the $A^{\sin(\phi_2+\phi_S-\pi)}_{2CL}$ (blue points) vs
    $\Delta\phi$.
Right: the $A^{\cos(\phi_1+\phi_S-\pi)}_{1CL}$ (red points) and
    the $A^{\cos(\phi_2+\phi_S-\pi)}_{2CL}$ (blue points) vs
    $\Delta\phi$.
\label{fig:a1cl}}
\end{figure}

The analysis presented here
refers to the so called ``$2h$ sample'', namely SIDIS events 
where at least one pair of 
oppositely charged final state hadrons has been reconstructed. 
The selection of such events is described in detail in 
Ref. \refcite{Adolph:2014fjw}. 
Specific to this analysis is the requirement that 
a minimum value of 0.1 GeV/c for the hadron transverse momenta 
 $p_{T \,(1,2)}^h$  (the indexes 1 and 2 refer to the positive and negative 
hadron of the pair respectively) is required to ensure good resolution in the 
azimuthal 
angles $\phi_1$ and $\phi_2$ measured in 
the standard gamma-nucleon reference system for 
SIDIS \cite{Bacchetta:2004jz}. 
Also, the two hadrons must 
carry a fraction $z_{1,2}$ of the virtual-photon energy of at least 0.1
and the results presented here have been obtained in the region
$x>0.032$.
For the positive and negative hadrons in the $2h$ sample the Collins-like
asymmetries have been extracted, namely the amplitudes 
$A^{\sin(\phi_1+\phi_S-\pi)}_{1CL}$
and $A^{\sin(\phi_2+\phi_S-\pi)}_{2CL}$ of the $\sin(\phi_{1,2} + \phi_S - \pi)$ 
modulations in the 
cross section, 
where $ \phi_S$  is the azimuthal angle of the transverse spin of the target 
proton.
For this analysis charged hadrons are used as many times as 
in the hadron pairs. 
To better investigate the mirror symmetry and the anti-correlation between
$\phi_1$ and $\phi_2$ the Collins asymmetries have been measured
as functions of the angle $\Delta \phi = \phi_1-\phi_2$. 
The results are shown in the left panel of
Fig.~\ref{fig:a1cl} .
The two asymmetries look like even functions of $\Delta \phi$,
are compatible with zero 
when $\Delta \phi$ tends to zero, and increase in magnitude 
as $\Delta \phi$ increases. 
The mirror symmetry between positive and negative hadrons is again a 
striking feature of the data. 

The trend of the data can be described rather well starting from the 
general expression for the transversity induced part of 
the SIDIS cross-section of 
Ref.~\refcite{Kotzinian:2014uya}.
After integration on $x$, $Q^2$, $z_1$, $z_2$, $p^2_{T1}$ and $p^2_{T2}$, 
the cross-section for the SIDIS process $lN \rightarrow l^{'} h^+h^- X$
can be
written as
\begin{equation}
\label{eq:eq_asy_1}
\frac{d\sigma^{h_1h_2}}{d\phi_1 d\phi_2 d\phi_S}
 = \sigma^{h_1h_2}_{U} + S_T \left[
\sigma^{h_1h_2}_{1C}\sin(\phi_1+\phi_S-\pi)  +
\sigma^{h_1h_2}_{2C}\sin(\phi_2+\phi_S-\pi)
\right], 
\end{equation}
where the  structure functions 
$\sigma^{h_1h_2}_{1C}$, $\sigma^{h_1h_2}_{2C}$ and $\sigma^{h_1h_2}_{U}$ 
might depend on $\Delta \phi$.
It is interesting to rewrite
Eq.~(\ref{eq:eq_asy_1}) 
 in terms of $\phi_1$ and $\Delta \phi$, or alternatively in terms of
$\phi_2$ and $\Delta \phi$:
\begin{equation}
\label{eq:eq00}
\begin{split}
\frac{d \sigma^{h_1h_2} }{d\phi_1 d \Delta\phi d\phi_S} =  \sigma_U^{h_1h_2} + S_T  & \Bigl[ \left( \sigma_{1C}^{h_1h_2}+\sigma_{2C}^{h_1h_2}\cos \Delta\phi \right) \sin(\phi_1+\phi_S-\pi) \Bigr.. \\
                                                                                 & \Bigl. - \sigma_{2C}^{h_1h_2} \sin \Delta\phi \cos(\phi_1+\phi_S-\pi) \Bigr], \\
\frac{d \sigma^{h_1h_2}}{d\phi_2 d \Delta\phi d\phi_S}   =  \sigma_U^{h_1h_2} +S_T  & \Bigl[ \left( \sigma_{2C}^{h_1h_2}+\sigma_{1C}^{h_1h_2}\cos \Delta\phi \right) \sin(\phi_2+\phi_S-\pi) \Bigr. \\
                                                                                 & \Bigl.  + \sigma_{1C}^{h_1h_2} \sin \Delta\phi \cos(\phi_2+\phi_S-\pi) \Bigr] . 
\end{split}
\end{equation}
With the change of variables above a new 
modulation, of the type $\cos(\phi_{1,2}+\phi_S-\pi)$, appears in the 
cross section, which can then be rewritten alternatively in terms of 
the sin and cos modulations of only one of the two hadrons. 
Fig.~\ref{fig:a1cl} (right panel) shows the amplitudes 
$A_{1,2CL}^{\cos(\phi_{1,2}+\phi_S-\pi)}$
of the
$\cos(\phi_{1.2}+\phi_S-\pi)$ modulations
obtained from our data: 
the amplitudes are rather similar for  positive and negative hadrons, 
and they seem to be odd-function of $\Delta\phi$. 
 
\begin{figure}[bt]
  \centering
  \includegraphics[width=0.475\textwidth]{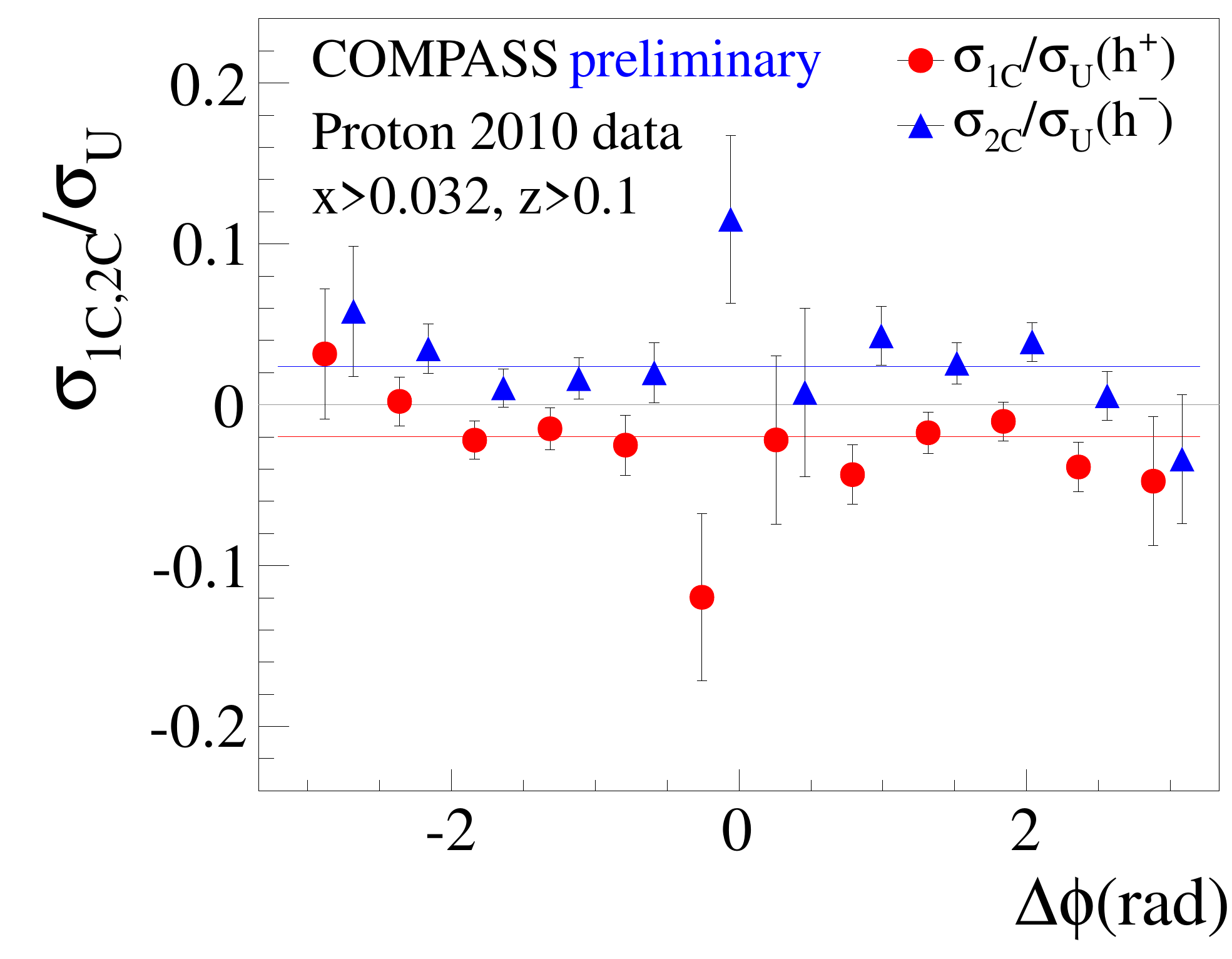}
\caption{$\sigma^{h_1h_2}_{1C} / \sigma^{h_1h_2}_{U}$ 
and $\sigma^{h_1h_2}_{2C} /
\sigma^{h_1h_2}_{U}$ as extracted from the measured asymmetries.
 \label{fig:rcs}}
\end{figure}

From Eq.~(\ref{eq:eq00}) it is possible to write explicit expressions 
for the four amplitudes shown in Fig.~\ref{fig:a1cl}:
\begin{eqnarray}
  \label{eq:eq1}
  A_{1CL}^{\sin(\phi_1+\phi_S-\pi)}(\Delta \phi) &=& 
  \frac{\sigma_{1C}^{\sin(\phi_1+\phi_S-\pi)}}{\sigma_U^{h_1h_2}}
  =\frac{\sigma_{1C}^{h_1h_2}(\Delta \phi)
    +\sigma_{2C}^{h_1h_2}(\Delta \phi) \cos \Delta \phi}{\sigma_U^{h_1h_2}(\Delta \phi)} \nonumber \\
  A_{1CL}^{\cos(\phi_1+\phi_S-\pi)}(\Delta \phi) &=& 
  \frac{\sigma_{1C}^{\cos(\phi_1+\phi_S-\pi)}}{\sigma_U^{h_1h_2}}
  =\frac{-\sigma_{2C}^{h_1h_2}(\Delta \phi)
    \sin \Delta \phi}{\sigma_U^{h_1h_2}(\Delta \phi)} \nonumber \\
  A_{2CL}^{\sin(\phi_2+\phi_S-\pi)}(\Delta \phi) &=& 
  \frac{\sigma_{2C}^{\sin(\phi_2+\phi_S-\pi)}}{\sigma_U^{h_1h_2}}
  =\frac{\sigma_{2C}^{h_1h_2}(\Delta \phi)
    +\sigma_{1C}^{h_1h_2}(\Delta \phi) \cos \Delta \phi}{\sigma_U^{h_1h_2}(\Delta \phi)} \nonumber \\
  A_{2CL}^{\cos(\phi_2+\phi_S-\pi)}(\Delta \phi) &=& 
  \frac{\sigma_{2C}^{\cos(\phi_2+\phi_S-\pi)}}{\sigma_U^{h_1h_2}}
  =\frac{\sigma_{1C}^{h_1h_2}(\Delta \phi)
    \sin \Delta \phi}{\sigma_U^{h_1h_2}(\Delta \phi)}
\end{eqnarray}
The quantities $\sigma^{h_1h_2}_{1} / \sigma^{h_1h_2}_{U}$ and $\sigma^{h_1h_2}_{2} /
\sigma^{h_1h_2}_{U}$, which
in principle can still be functions of $\Delta \phi$, 
can  be obtained from the measured asymmetries
since it is:
\begin{eqnarray}
\label{eq:eq3}
\frac{\sigma_{1C}^{h_1h_2}(\Delta \phi)}{\sigma_U^{h_1h_2}(\Delta \phi)}
&=& 
A_{1CL}^{\sin (\phi_1+\phi_S-\pi)}(\Delta \phi) + 
A_{1CL}^{\cos (\phi_1+\phi_S-\pi)}(\Delta \phi) \cot \Delta \phi
 \nonumber \\
\frac{\sigma_{2C}^{h_1h_2}(\Delta \phi)}{\sigma_U^{h_1h_2}(\Delta \phi)}
&=& 
A_{2CL}^{\sin(\phi_1+\phi_S-\pi)}(\Delta \phi) -
A_{2CL}^{\cos (\phi_2+\phi_S-\pi)}(\Delta \phi) \cot \Delta \phi
\end{eqnarray}
The results for the ratios of the cross-sections are given in 
Fig.~\ref{fig:rcs} and show that 
within statistical errors
$\sigma^{h_1h_2}_{1C} /\sigma^{h_1h_2}_{U}$  and 
$\sigma^{h_1h_2}_{2C} /\sigma^{h_1h_2}_{U}$ are constant, equal in absolute value 
and of opposite sign.
\begin{figure}[bt]
  \centering
  \includegraphics[width=0.95\textwidth]{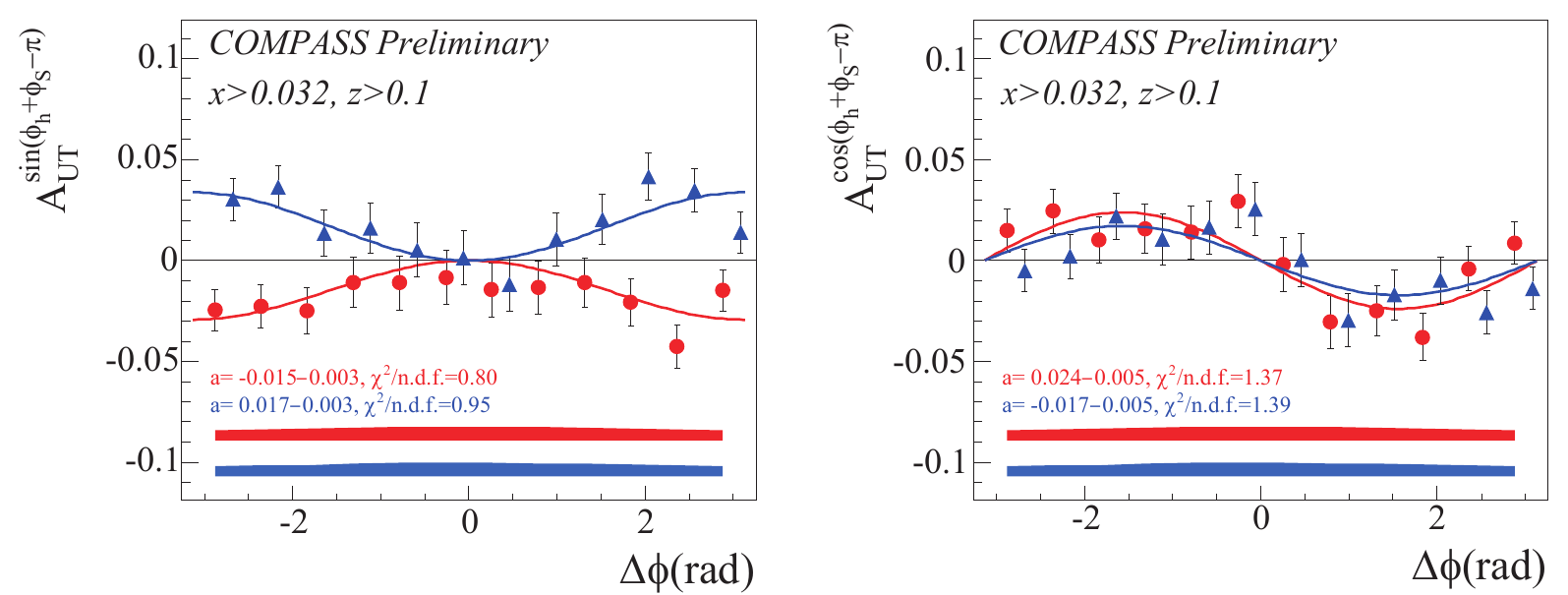}
\caption{Same data points as in Fig.~\ref{fig:a1cl}
with superimposed the fitting functions (see text).}
\label{fig:fxx}
\end{figure}
As a consequence, the measured asymmetries can well be fitted
 with
the simple functions
$\pm a \cdot (1-\cos\Delta\phi)$  in the case of the sine asymmetries
(Fig.~\ref{fig:fxx}, left), and
$a \sin\Delta\phi$ for the cosine asymmetries (Fig.~\ref{fig:fxx}, right),
with values for the constant $a$ well compatible for
the four asymmetries.
As a conclusion, the mirror symmetry between the positive and negative 
hadron asymmetries is clearly confirmed from these studies
and can be summarized by the relation $\sigma^{h_1h_2}_{1C} /\sigma^{h_1h_2}_{U}
= - \sigma^{h_1h_2}_{2C} /\sigma^{h_1h_2}_{U}$ where the ratios
of the structure functions do not depend on $\Delta\phi$.

In Refs.~\refcite{fb_dubna,Adolph:2014fjw} it was shown that the 
Collins asymmetry of the hadron pair, defined as the amplitude of the 
modulation $\sin(\phi_{2h}+\phi_S-\pi)$, where 
$\phi_{2h} = [\phi_1 + (\phi_2 - \pi)] / 2 $
(calculated using the unit vectors of the transverse nomenta),
is essentially identical 
to the dihadron asymmetry which is calculated in the standard
procedure~\cite{Artru:2002pua,Adolph:2014fjw}
which uses the transverse component of the relative momentum between the 
two hadrons. 
Starting from the general expression of the cross section given
in Eq.~(\ref{eq:eq_asy_1}), it is possible to calculate the
amplitude of the modulation  $\sin(\phi_{2h}+\phi_S-\pi)$.
Changing variables from $\phi_1 $ and $\phi_2$ to $\phi_1$ and $\phi_{2h}$ 
and using
$\sigma^{h1h2}_{2C} = - \sigma^{h1h2}_{1C}$, Eq. (\ref{eq:eq_asy_1})
can be rewritten as:
\begin{eqnarray}
\sigma^{h1h2} = \sigma^{h1h2}_{U} + S_T \cdot \sigma^{h1h2}_{1C} \cdot
\sqrt{2(1-\cos \Delta\phi )}\cdot \sin(\phi_{2h}+\phi_S -\pi).
\label{eq:s2h}
\end{eqnarray}

\begin{figure}[tb]
  \centering
  \includegraphics[width=0.95\textwidth]{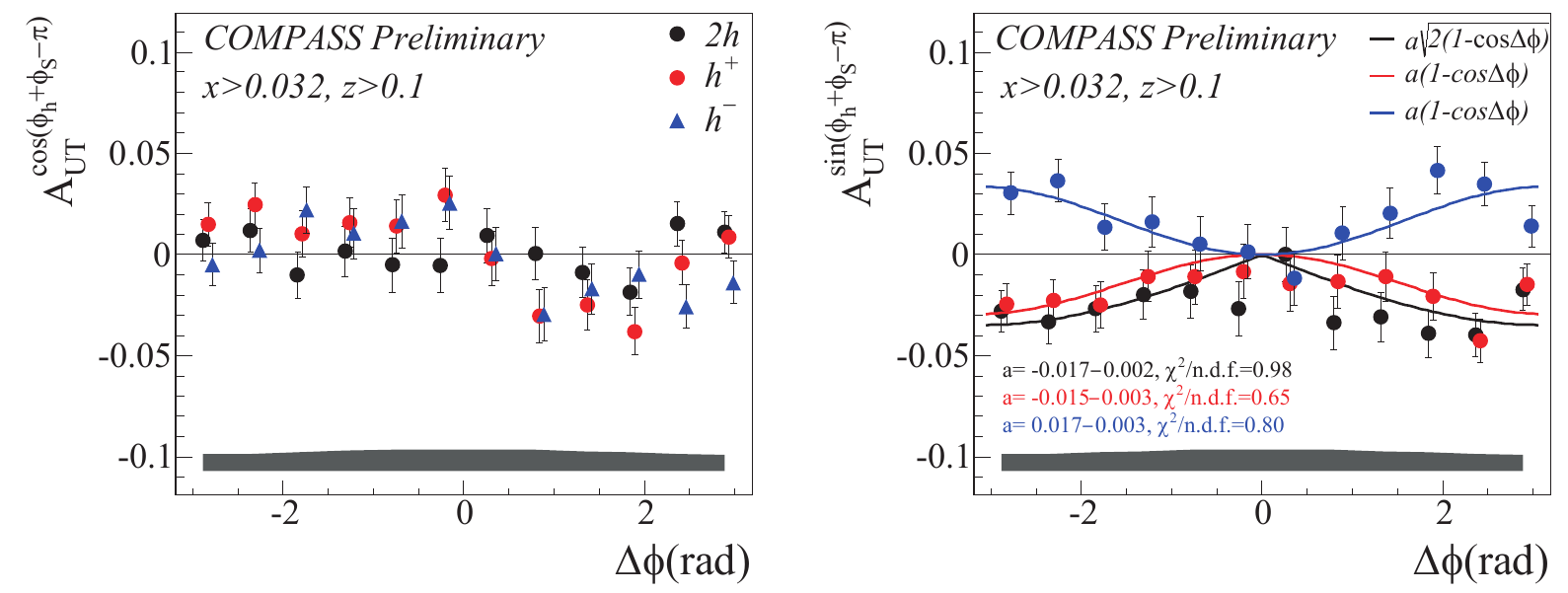}
  \caption{Left: $A^{\cos(\phi_1+\phi_S-\pi)}_{1CL}$ (red points), 
 $A^{\cos(\phi_2+\phi_S-\pi)}_{2CL}$ (blue points) and  
$A^{\cos(\phi_{2h}+\phi_S-\pi)}_{2h,CL}$ (black points) vs
$\Delta\phi$. 
Right:  $A^{\sin(\phi_1+\phi_S-\pi)}_{1CL}$ (red points), 
 $A^{\sin(\phi_2+\phi_S-\pi)}_{2CL}$ (blue points) and  
$A^{\sin(\phi_{2h}+\phi_S-\pi)}_{2h,CL}$ (black points) vs
$\Delta\phi$.\label{fig:a12h}}
  \end{figure}

This cross section implies a sine modulation with the amplitude
\begin{eqnarray}
A^{\sin(\phi_{2h}+\phi_S-\pi)}_{2h,CL} = \frac{
  \sigma^{h1h2}_{1C}(\Delta\phi)}{\sigma^{h1h2}_{U}(\Delta\phi)} \cdot
\sqrt{2(1-\cos \Delta\phi)}
\end{eqnarray}
while no $A^{\cos(\phi_{2h}+\phi_S-\pi)}_{2h,CL}$ asymmetry is expected.
Our measurements are shown in Fig. \ref{fig:a12h}.
As can be seen in the left panel, the 
$A^{\cos(\phi_{2h}+\phi_S-\pi)}_{2h,CL}$ asymmetry (black points) is
indeed compatible with zero.
In the right panel of 
Fig.~\ref{fig:a12h} the $A^{\sin(\phi_{2h}+\phi_S-\pi)}_{2h,CL}$, 
the $A^{\sin(\phi_1+\phi_S-\pi)}_{1CL}$
and the $A^{\sin(\phi_2+\phi_S-\pi)}_{2CL}$ asymmetries are shown together with the 
 curves $p_{0,1(2)}\cdot(1-\cos(\Delta\phi))$ (red, blue lines) and
$p_{0,2h}\cdot\sqrt{2(1-\cos(\Delta\phi))}$ (black line) 
as obtained by the fits.
As can be seen the fits are very good, and the values of the $p_{0x}$
parameters
are all compatible, in agreement with the fact that
$ \sigma^{h1h2}_{1C} / \sigma^{h1h2}_{U}$
is the same for the three asymmetries.
Evaluating 
the ratio of the integrals 
of the two-hadron amplitudes over the one-hadron amplitudes
one gets a value of
$4/\pi$ which is in agreement with our original observation
that the dihadron asymmetry is somewhat larger than the Collins asymmetry
for positive hadrons.

As a conclusion, we have shown that in the SIDIS process the Collins 
asymmetries of the positive and negative hadrons are mirror symmetric, 
in agreement with a  $(1-\cos \Delta \phi)$  dependence 
which is expected from general principles. 
Most important, the amplitude of the dihadron asymmetry has a 
very simple relation to that of the single hadron asymmetries, 
confirming our original observation that the Collins mechanism and 
the interference fragmentation must have a common physical origin.   

A similar analysis has been performed for the Sivers asymmetry,
namely for modulations of the type $\sin(\phi_{1,2,2h} - \phi_S)$,
but as expected from the $h^+$ and $h^-$
data~\cite{Adolph:2012sp} 
and from model calculations~\cite{Kotzinian:2014uya}
no mirror symmetry is observed for
positive and negative hadrons, and no simple relationship is extracted for the
structure functions.


\end{document}